# Games and Culture: Using Online-gaming Data to Cluster Chinese Regional Cultures

Xianwen Wang[1,2,3]*, Wenli Mao[1,2], Chen Liu[1,2]

[1]WISE Lab, Faculty of Humanities and Social Sciences, Dalian University of Technology, Dalian 116085, China.
[2]School of Public Administration and Law, Dalian University of Technology, Dalian 116085, China.
[3]DUT- Drexel Joint Institute for the Study of Knowledge Visualization and Scientific Discovery, Dalian University of Technology, Dalian 116085, China.

* Corresponding author.
Email address: xianwenwang@dlut.edu.cn

**Abstract**
To identify cluster of societies and cultures is not easy in subject to the availability of data. In this study, we propose a novel method to cluster Chinese regional cultures. Using geotagged online-gaming data of Chinese internet users playing online card and board games with regional features, 336 Chinese cities are grouped into 17 clusters. The distribution of clustering units shows great geographical proximity when the boundary of the clusters coincides well with the geographical boundary of provinces.

**Introduction**

Unlike the massive data available for natural sciences, data in social science are hard to collect (Latour, 2007). However, with the vast geotagged usage data generated by people in the web, it is possible for researchers to reveal and understand details of both individual and social behavior with unprecedented detail (Manovich, 2011; Girardin et al., 2008; Eagle and Pentland, 2006). Based on the fast development of big data, it's possible for researchers to create and define new methods of observing, recording, and analyzing human dynamics (O'Neill et al., 2006).

With the data collected from 100 Bluetooth-enabled mobile phone users for 9 months, Nathan Eagle & Pentland make a good try to capture daily human behavior, "identify socially significant locations, and model organizational rhythms" (Eagle and Pentland, 2006). Using the geotagged usage data of scientific paper downloading from Springer Verlag, we try to capture details of researchers' daily working behavior (Wang et al., 2012).

To identify cluster of societies is not easy in subject to the availability of data. There has been much effort to group countries into similar clusters using survey data (Cattell, 1950; Gupta et al., 2002; Smith et al., 1996; Brodbeck et al., 2000) (Furnham et al., 1994). Previous studies have shown that many factors, i.e. religious and linguistic commonality, geographic proximity, and mass migrations and ethnic social capital are relevant factors in the clustering of societies (Gupta et al., 2002).

As one of the world's earliest civilization, most populous and second largest country by land area, China has diversified cultures. i.e. China has as many as 292 living languages, people in different cities usually can't standard each other if they speak dialect, especially in south China. For a long time, Chinese culture are roughly divided into two parts, south and north, along the Qinling

Mountains-Huaihe River line. There are some other classifications, including three zones (the eastern, central, and western region), six economic zones (northeast, eastern, central and western economic region), etc. However, these classifications are rather rough and lack of evidence.

In this study, using a kind of novel large sample usage data, we try to cluster Chinese cultures from a new perspective.

**Data Source and Preparation**
*QQ Game Map*
QQ is an abbreviation of Tencent QQ, which had been called OICQ during the period of 1998 – 2000. Now QQ is the most popular instant messenger software in China. By the end of 2012, there were 798 million active user accounts with approximately 170 million users online at a time.

QQ Game is a casual games client, offering only multi-player online games. 196 board and card games are available through the client. There are approximately 8 million active online players during peak hours, generally 21:00 – 23:00. Online players once peaked to 9.4 million in December 2012.

QQ Game Map is a service launched by Tencent, Inc. since October 22, 2012. It provides real-time visualization map of geographical distribution QQ game online players. Besides general distribution of the total QQ game players, the map also provide visualization for any single QQ game.

*Data Preparation*
Our data is collected from the website of QQ Game Map (http://qqgame.qq.com/online.shtml). For the 196 kinds of QQ games and 376 cities in China, the number of online players of each game in each city is recorded at the time 21:45, August 9, 2013, as table 1 shows.

**Table 1.** Data collecting

| City | Total players | Shanghai mahjong | Sichuan mahjong | … |
|---|---|---|---|---|
| Shanghai | 205,568 | 8740 | 1078 | … |
| Beijing | 235,185 | 100 | 1775 | |
| Chongqing | 134,675 | 5 | 11,312 | … |
| Shenzhen | 169,112 | 17 | 770 | … |
| … | … | … | … | … |
| All cities | 6,536,549 | 14,260 | 75,978 | … |

**Results**
*Geographical distribution*
Figure 1 shows the geographical distribution of QQ game players. The node size is correlated to the number of players of the city. The white shining spot in the center of the node means that someone of the region is beginning to play game.

There are approximate 7.55 million players online at this time point. Most nodes are clustered together in the east coastal regions, north China plain (including Beijing, Tianjin and Shandong province), when the dark west China form a powerful contrast to the dazzling east regions.

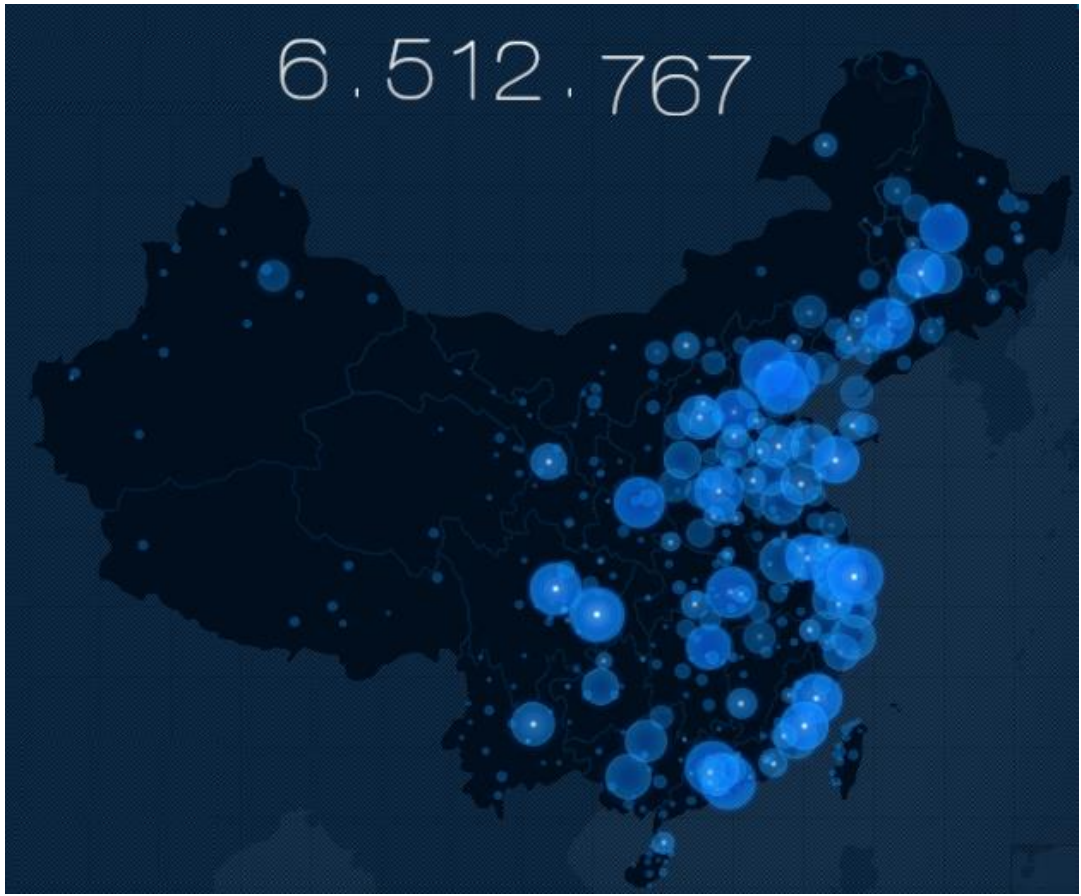

**Figure 1.** Screenshot of QQ Game Map

Beijing is the most active city in China in terms of online game playing, with over 0.23 million players online at the time, which accounts for about 1.20% of its population. Shanghai ranks number 2 with slightly less. The number 3 city is Shenzhen with approximately 0.17 million players and a high player-to-population ratio (1.63%). Shenzhen is the youngest big city located in the south of China and is adjacent to Hong Kong, which owns a number of factories and armies of young migrant workers. It is not strange that Shenzhen has the highest player-to-population ratio because of the plenty of young people aged from 18 to 40.

In Table 2, we list the top 20 cities with most QQ game players at the time of 21:45, August 9, 2013. The total number of the 20 cities is about 2.15 million and accounts for 31.89% of all QQ game players in the whole China.

**Table 2.** Top 20 cities with most QQ game players

| Rank | City | Province | Number of players | Population | Percent |
|------|------|----------|-------------------|------------|---------|
| 1 | Beijing | NA | 235,185 | 19,612,368 | 1.20% |
| 2 | Shanghai | NA | 205,568 | 23,019,148 | 0.89% |
| 3 | Shenzhen | Guangdong | 169,112 | 10,357,938 | 1.63% |
| 4 | Chongqing | NA | 134,675 | 28,846,170 | 0.47% |
| 5 | Taiyuan | Shanxi | 127,579 | 4,201,591 | 3.04% |
| 6 | Guangzhou | Guangdong | 123,010 | 12,700,800 | 0.97% |
| 7 | Xi'an | Shaanxi | 120,222 | 8,467,837 | 1.42% |
| 8 | Hangzhou | Zhejiang | 111,637 | 8,700,400 | 1.28% |
| 9 | Chengdu | Sichuan | 107,915 | 14,047,625 | 0.77% |
| 10 | Tianjin | NA | 100,430 | 12,938,224 | 0.78% |
| 11 | Suzhou | Jiangsu | 88,091 | 10,465,994 | 0.84% |
| 12 | Wuhan | Hubei | 85,028 | 9,785,392 | 0.87% |
| 13 | Dongguan | Guangdong | 74,208 | 8,220,237 | 0.90% |
| 14 | Ningbo | Zhejiang | 70,878 | 7,605,689 | 0.93% |
| 15 | Fuzhou | Fujian | 70,068 | 7,115,370 | 0.98% |
| 16 | Quanzhou | Fujian | 68,601 | 8,128,530 | 0.84% |
| 17 | Zhengzhou | Henan | 67,710 | 8,626,505 | 0.78% |
| 18 | Jilin City | Jilin | 64,847 | 4,414,681 | 1.47% |
| 19 | Fuyang | Anhui | 63,157 | 7,599,918 | 0.83% |
| 20 | Shijiazhuang | Hebei | 62,064 | 10,163,788 | 0.61% |
| | All top 20 | NA | 2,149,985 | 225,018,205 | 0.96% |
| | All China | NA | 6,536,549 | 1,339,724,852 | 0.49% |

Note: population data are collected from *Communiqué of the National Bureau of Statistics of People's Republic of China on Major Figures of the 2010 Population Census (No. 2)*

*Regional difference and cultural diversity in China*

Playing card and board game is a favorite pastime for Chinese people. China has strong board and card game culture. There are hundreds of card and board games in China. Most provinces and many cities have distinct card and board games. For example, there are more than 30 kinds of mahjong in China. Shanghai has Shanghai mahjong. Zhejiang, a neighbor province of Shanghai, has Hangzhou mahjong and Ningbo mahjong. So is the case for card games.

Figure 2 shows the geographical distribution of players of 13 Chinese mahjong games, including Sichuan mahjong, Hangzhou mahjong, Wuhan mahjong, Harbin mahjong, etc. Each mahjong game has its own sphere of influence, there may be some overlap between two games, but generally the boundary of games is distinct.

For Hangzhou mahjong, as the red dots shows, they are concentrated in Zhejiang province and other Yangtze River delta regions, including Shanghai and Jiangsu province.

For another kind of mahjong game, Sichuan mahjong, most players are concentrated in the west China, including Sichuan province, Chongqing city, Guizhou province, Yunnan province. Other regions, e.g. Yangtze River delta in east China, Pearl River Delta in south China, also have scattered distribution.

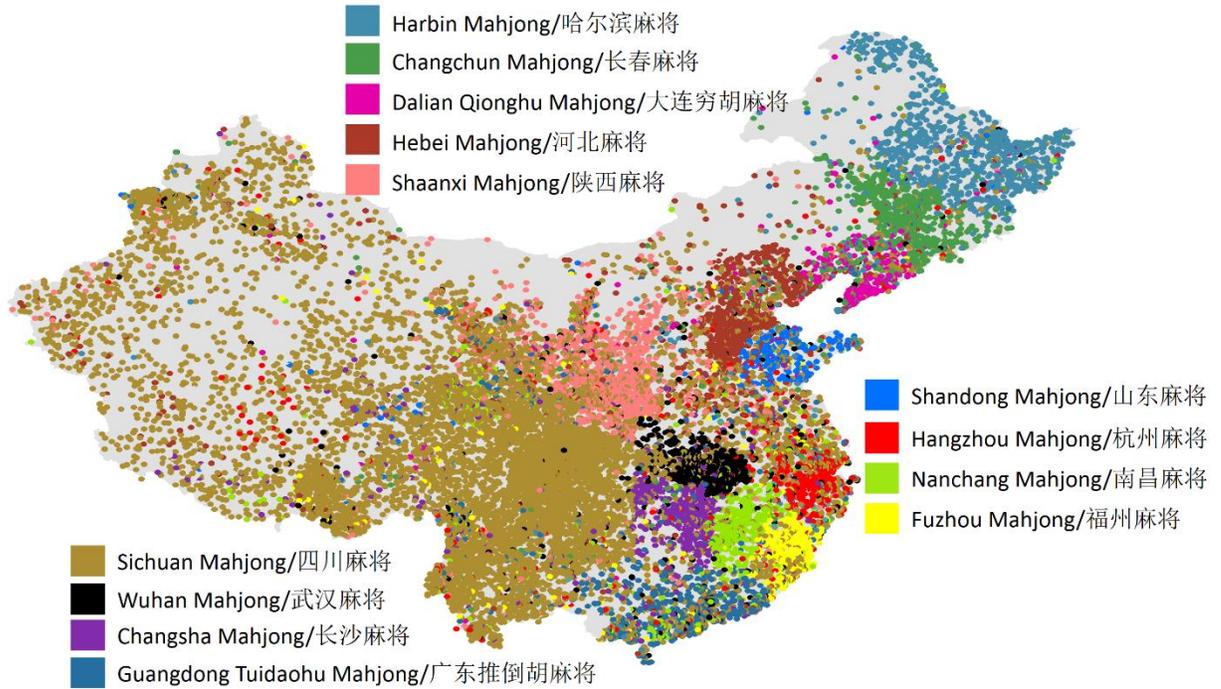

**Figure 2.** Geographical distribution of mahjong players with distinct regional characteristics

For contrast, we also choose another two games as control sample, which are Tractor and Happy Bullfight. They are also popular games in China and without much regional feature. As Figure 3 shows, the nodes distribution is very consistent with the general distribution shown in Figure 1.

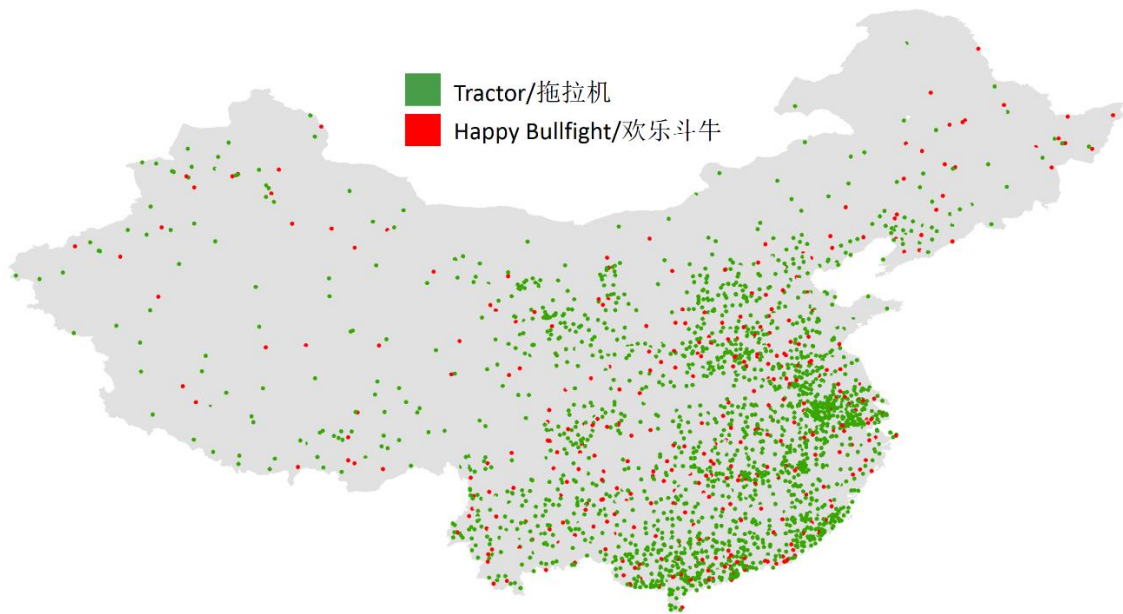

**Figure 3.** Geographical distribution of games without regional characteristics

So, in order to identify the interregional similarities and interregional differences, only games with distinct regional features are selected from the complete dataset of 196 games. The criterion for regional game is quantified by the following rule, players from the top 20 cities account for over 50% of all players, when players from the top 5 cities account for less than 70%. Then, we make correlation analysis for the games. If games are high correlated with each other, only one (with more players) could be kept from the highly correlated games (correlation coefficient greater than 0.8).

For example, as shown in Table 3, the proportion of players in top 5 cities for the three games, Tractor and Sichuan Mahjong are 18.76% and 40.04% respectively, when the proportion of Shanghai Mahjong is as high as 90.42%. According to the criterion mentioned above, Shanghai Mahjong is excluded. When considering the proportion of players in top 20 cities, Tractor is excluded because its proportion is less than 50%.

Table 3. Percentage of players of top 5/20 cities in total

|  | Tractor | Sichuan Mahjong | Shanghai Mahjong |
|---|---|---|---|
| % of top 5 cities | 18.76% | 40.04% | 90.42% |
| % of top 20 cities | 43.16% | 60.32% | 96.45% |

Figure 4 shows the distribution of 3 games. As a kind of widespread game in China, Tractor has little regional characteristic, which could be seen from the smooth distribution of blue dots. When the distribution of Shanghai Mahjong, illustrated with red dots, is very steep. It means that players of Shanghai Mahjong concentrated in few cities, i.e. Shanghai. And the distribution of Sichuan Mahjong is somehow between.

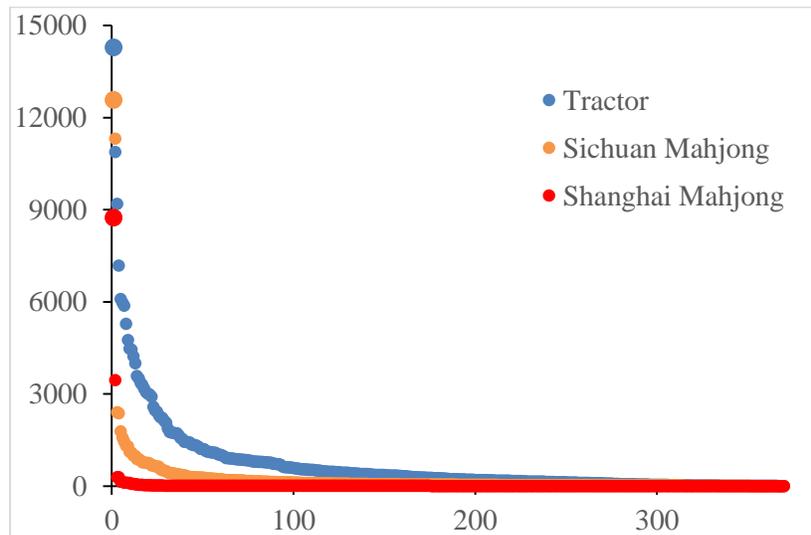

Figure 4. The distribution of 3 QQ games

Finally, 47 games with distinct regional characteristic are selected. For each provincial region in China, 2-3 kinds of QQ games with regional features are included as research samples, as Table 4 shows.

Table 4. 47 QQ games included in this study

| Rank | Name | Chinese name | No. of players | Rank | Name | Chinese name | No. of players |
|---|---|---|---|---|---|---|---|
| 1 | Chinese Poker | 四人斗地主 | 195485 | 25 | Baofen | 包分 | 6299 |
| 2 | Baohuang | 保皇 | 78146 | 26 | 4A4 | 4A4 | 6239 |
| 3 | Sichuan Mahjong | 四川麻将 | 75978 | 27 | Atom | 原子 | 5661 |
| 4 | Yaodiren | 么地人 | 59681 | 28 | Guanpai | 关牌 | 5261 |
| 5 | Protean Shuangkou | 千变双扣 | 39755 | 29 | Erqiwang | 二七王 | 5163 |
| 6 | Da Da A | 打大 A | 32521 | 30 | Nanchang Mahjong | 南昌麻将 | 5116 |
| 7 | New Gouji | 新够级 | 29163 | 31 | Dalian Qionghu Mahjong | 大连穷胡麻将 | 4854 |
| 8 | Dig | 挖坑 | 26296 | 32 | Changsha Mahjong | 长沙麻将 | 4487 |
| 9 | Chinese Hearts | 拱猪 | 23084 | 33 | Tianjin Mahjong | 天津麻将 | 3240 |
| 10 | Guandan | 掼蛋 | 21962 | 34 | Shandong Mahjong | 山东麻将 | 2848 |
| 11 | Paohuzi | 跑胡子 | 16426 | 35 | Harbin Mahjong | 哈尔滨麻将 | 2729 |
| 12 | Wuhan Mahjong | 武汉麻将 | 13280 | 36 | 240 | 二百四 | 2687 |
| 13 | Bieqi | 憋七 | 12759 | 37 | Situan | 四团 | 2671 |
| 14 | Hebei Mahjong | 河北麻将 | 12124 | 38 | Four Dig | 四人挖坑 | 2397 |
| 15 | Big Two | 锄大地 | 10435 | 39 | Black A | 黑尖 | 2280 |
| 16 | Sandayi | 三打一 | 10369 | 40 | Shoubayi | 手把一 | 2131 |
| 17 | Red 10 | 红十 | 9869 | 41 | Fuzhou Mahjong | 福州麻将 | 1893 |
| 18 | Sandaha | 三打哈 | 9162 | 42 | Changchun Mahjong | 长春麻将 | 1863 |
| 19 | Hangzhou Mahjong | 杭州麻将 | 8695 | 43 | Xinyang Black 7 | 信阳黑七 | 1741 |
| 20 | Shaanxi Mahjong | 陕西麻将 | 7642 | 44 | Paoyao | 刨幺 | 1560 |
| 21 | Guangdong Tuidaohu Mahjong | 广东推倒胡麻将 | 6991 | 45 | Guiyang Zhuoji Mahjong | 贵阳捉鸡麻将 | 946 |
| 22 | Guangdong Jipinghu Mahjong | 广东鸡平胡麻将 | 6685 | 46 | Wenzhou Mahjong | 温州麻将 | 573 |
| 23 | Ningbo Mahjong | 宁波麻将 | 6679 | 47 | Huashui Mahjong | 滑水麻将 | 408 |
| 24 | Qingdao Baohuang | 青岛保皇 | 6474 | | | | |

*Geographical Region Clustering*

For the 376 cities in China, we record the number of players of each game in each city. Considering the population size of cities, we divide the number of players of one game by the total players of all games in the city. Because of the data missing of some regions in Hainan province, and the special status of Taiwan, Hong Kong and Macau, these areas are excluded. Finally, 336 cities in Chinese mainland are selected as research objects.

Hierarchical cluster analysis is employed to cluster the 336 cities. The clustering method is Average Linkage (Between Groups), when the distance is measured by Phi-square. To better illustrate the clusters and their relationship, we group some cities into one cluster according to the hierarchical cluster result. As Figure 5 shows, the above panel is the cluster result, there are 17 clusters. The color of clusters are different, we use the color ramp to label the relationship of clusters. Clusters with the more similar color have the more close relationship. The cities are projected to the geographical map of China with different color according to the clustering result as Figure 5 (a) shows. Cities in the same cluster are labeled with the same color. The gray line in the map illustrate the boundaries of provinces.

As Figure 5 (b) shows, firstly, the distribution of clustering units shows great geographical proximity. Although we make cluster analysis of the cities with not any spatial constraint to limit group membership to contiguous features, cities within the same cluster have typical geographical proximity, which could be demonstrated by large areas with the same color. Moreover, this kind of geographical proximity also applies to adjacent clusters, i.e. cluster 1 and 2 are located adjacent, the same for cluster 7, 8 and 9, cluster 13, 14 and 15.

Secondly, in general the boundary of the clusters coincides well with the geographical boundary of provinces. Some clusters are restricted in one province, i.e. cluster 17 and Shandong province, cluster 16 and Hunan province. Some clusters are concentrated in two or more provinces. For example, cluster 1 includes Shanghai, Jiangsu and Anhui. Cluster 5 is consist of Jiangxi, Guangdong and Guangxi. Cluster 7 contains Shaanxi, most areas in Ningxia and Gansu and part of Qinghai.

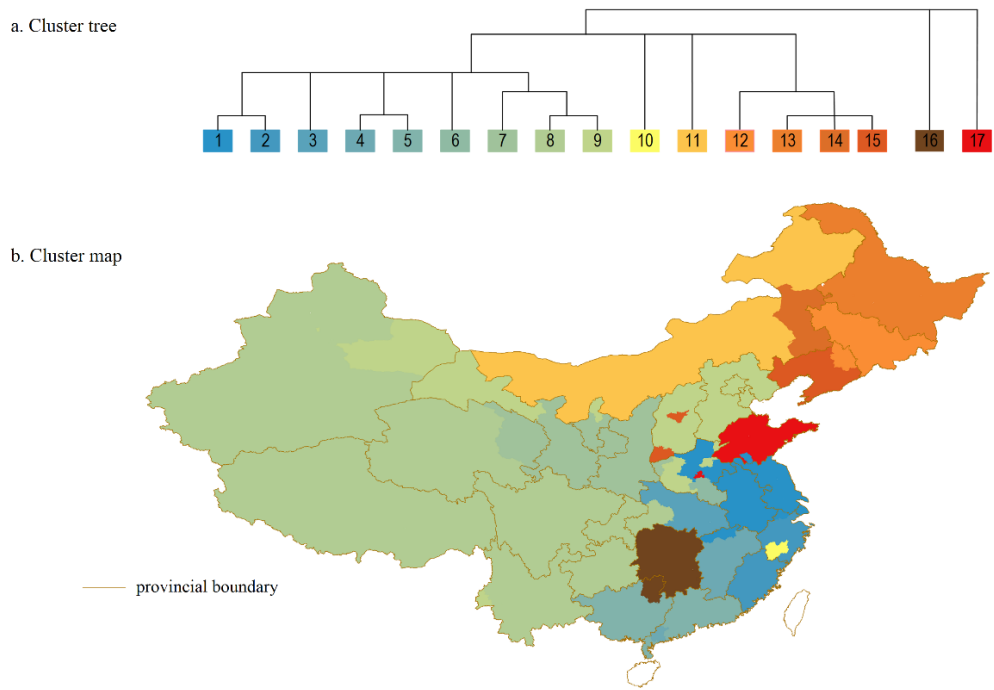

**Figure 5.** Result of cluster analysis

**Discussion**

Playing card and board game has a long history in Chinese societies, which has distinct cultural and societal characteristics. Using geotagged online-gaming data of Chinese internet users, we cluster Chinese regions into 17 groups. As the clustering results show, although we make cluster analysis of the cities with not any spatial constraint to limit group membership to contiguous features, cities within the same cluster have typical geographical proximity. The geographic boundaries of clusters coincide well with the boundaries of provincial regions, which indicate that regions in the same province tend to have similar cultures.

**Declaration of Conflicting Interests**

The authors declared no potential conflicts of interest with respect to the research, authorship, and/or publication of this article.

**Funding**

The work was supported by the project of "National Natural Science Foundation of China" (61301227); and the project of "Fundamental Research Funds for the Central Universities" (grant number DUT12RW309).